\documentclass[aps,prl,twocolumn,groupedaddress,floatfix]{revtex4}

\usepackage{graphicx}
\usepackage{subfigure}
\usepackage{epsfig}
\usepackage{amsmath}
\usepackage{bbm}

\def\be{\begin{equation}}
\def\ee{\end{equation}}
\def\bea{\begin{eqnarray}}
\def\eea{\end{eqnarray}}

\begin{document}

\title{Auger transitions in one-dimensional metals}

\author{E. Perfetto}

\affiliation{Dipartimento di Fisica Universita' di Roma Tor
Vergata, Via della Ricerca Scientifica 1, 00133 Roma, Italy.}

\affiliation{}

\date{\today}

\begin{abstract}

We present a dynamical theory of the Auger decay in
one-dimensional (1D) metals described by the Tomonaga-Luttinger
model. An analytic expression of the Auger current is derived in
the framework of the 1-step approach, where the finite lifetime of
the initial core-hole and the core-valence interaction are taken
into account. This allows to capture typical dynamical features
like the \textit{shake-down} effect, in which the Auger spectrum
shows a non-vanishing weight above the 2-step high-energy
threshold. The obtained results give also a hint to understand the
sizable suppression of Auger spectral weight closed to the Fermi
energy recently observed in carbon nanotubes with respect to
graphite.

\end{abstract}

\maketitle

\section{Introduction}

The dynamics of the core-valence-valence (CVV) Auger transitions
in strongly correlated solids has been extensively studied during
the last three decades\cite{cinirew}. However, despite the great
interest devoted to this problem, several aspects are still not
well understood. The theoretical calculation of the Auger spectrum
of correlated solids is a challenging task because, beside the
intrinsic difficulty to deal with a many-body interacting system,
the creation of the core-hole and the Auger process itself are in
principle coherent events, involving virtual Auger transitions and
incomplete relaxation phenomena\cite{gunn}. The complete
formulation of the theory describing the Auger decays has been
provided in 1980 by Gunnarsson and Sch\"{o}nhammer
(GS)\cite{gunn}. In the framework of the so-called 1-step
approach, they derived a general formula for the Auger current by
treating the decay of the initial core-hole to all orders.
Unfortunately such formulation cannot be cast in terms of Green's
functions and is of hard implementation for practical purposes. A
significant progress can be done within the 2-step approximation,
where the photoemission and the Auger decay are considered as
independent events. In this framework, Cini\cite{cini1} and
Sawatzky\cite{saw} (CS) proposed a simple and elegant theory able
to provide a quantitative understanding of the experimental Auger
spectra of transition metals with (almost) closed valence bands.
An advantage of such theory is that it also provides a practical
scheme to estimate the value of the screened interaction from the
experimental spectra\cite{bennett,lof,maiti}. This is particularly
useful to support LDA+$U$ calculations\cite{LdaU}. In the case of
open-band systems the CS approach breaks down and no reliable
theory is currently available. Very recently Seibold and
collaborators\cite{antibound} presented a theory of the dynamical
two-particle response function in the 2D Hubbard model based on
the time-dependent Gutzwiller approximation. Although important
effects are not treated there (e.g., the finite lifetime of the
core-hole and the interaction of the core-hole with the valence
electrons), the theory provides a novel tool to attack the
calculation of the Auger spectrum in correlated open-band systems.

In this paper we develop a dynamical theory of the CVV Auger
transitions in an ideal 1D metal. In the CVV Auger decay two holes
are left in the valence band after the X-ray photoemission of a
deep core-hole. Here assume that the valence electrons form the
so-called Luttinger Liquid (LL), described by the
Tomonaga-Luttinger model. We also allow for the interaction
between the core-hole and the valence electrons and introduce a
term responsible for the Auger transition, which destroys the
core-hole and creates the Auger electron together with the two
valence holes (and viceversa). The corresponding Auger current is
calculated analitically by using the bosonization and equations of
motion methods.

The paper is organized as follows. In Section II we describe in
detail the model Hamiltonian used in the present work. In Section
III we derive a closed analytic expression for the Auger current
in framework of the 1-step approach. In Section IV we discuss the
relevant features emerging form the obtained formula. In
particular we show that the theory exposed here is able to capture
some striking features recently observed in the Auger spectra of
carbon nanotubes. Finally, a brief summary and the the main
conclusions are drawn in Section V.

\section{The Model}

The LL is the prototype of interacting electrons confined in one
spatial dimension, characterized by striking phenomena such as the
so-called spin-charge separation and the power-law dependence of
observables in proximity of the Fermi
energy\cite{haldane,schulz,giamarchi,schonhammer,voit,meden}. In
$H_{\rm{Lutt}}$ the electrons have a linear dispersion relation
around positive (Right) and negative (Left) Fermi points and the
electron-electron (e-e) interactions act only between Right/Left
electron densities. The model is exactly solvable by means of the
bosonization technique, which allows to write the electron
Hamiltonian in terms of boson operators $b$'s\cite{gdsv}:
\begin{eqnarray}
H_{\rm{Lutt}}&=&\sum_{q \neq 0,\sigma}  \frac{v_F}{2} |q|
\Big{[}b^{\dagger}_{\sigma}(q) b_{\sigma}(q)+b_{\sigma}(q)
b^{\dagger}_{\sigma}(q)\Big{]} \, , \nonumber
\\
&+& \sum_{q \neq 0,\sigma} \frac{g_4}{4\pi} |q|
\Big{[}b^{\dagger}_{\sigma}(q)b_{\sigma}(q)+b_{\sigma}(q)b^{\dagger}_{\sigma}(q)
\nonumber \\
&+&
b^{\dagger}_{\sigma}(q)b_{-\sigma}(q)+b_{\sigma}(q)b^{\dagger}_{-\sigma}(q)\Big{]}
\, , \nonumber \\ &-& \sum_{q \neq 0,\sigma}
 \frac{g_2}{4 \pi} |q|
\Big{[}b^{\dagger}_{\sigma}(q)b^{\dagger}_{\sigma}(-q)+b_{\sigma}(q)b_{\sigma}(-q)
\nonumber\\
&+&
b^{\dagger}_{\sigma}(q)b^{\dagger}_{-\sigma}(-q)+b_{\sigma}(q)b_{-\sigma}(-q)
\Big{]} \, , \label{seconda}
\end{eqnarray}
where $\sigma=\uparrow,\downarrow$ is the spin index,
$[b_{\sigma}(q),b^{\dagger}_{\sigma '}(q')]=\delta_{\sigma,\sigma
'}\delta_{q,q'}$, $v_F$ is the Fermi velocity, $g_{4}$ is the
interaction parameter between Right-Right (positive $q$) and
Left-Left (negative $q$) electron densities, while $g_{2}$ is the
interaction parameter between Left-Right densities.

The key point of the bosonization is that it is possible to
express the fermion fields in terms of boson fields. Here it is
useful to introduce a chirality index $\nu =R,L$ to distinguish
between Right and Left electron modes. For instance the $\nu = R$
fermion field is given by:
\begin{eqnarray}
\psi_{\sigma,R}(x)=\frac{\eta_{R,\sigma}}{(2 \pi
\alpha)^{1/2}}e^{i \Phi_{\sigma,R}(x)} \, , \label{bos1}
\end{eqnarray}
where $\eta_{R,\sigma}$ is an anticommuting Klein factor and
\begin{eqnarray}
\Phi_{\sigma,R}(x) &=& \sum_{q>0} \left( \frac{2\pi}{qL}
\right)^{1/2} e^{-\alpha q/2} \Big{[}b_{\sigma}^{\dagger}(q)
e^{-iqx}+ b_{\sigma}(q) e^{iqx}\Big{]} \nonumber \\  &+&
\varphi_{0,R}+ 2\pi x N_R /L \, , \label{boson}
\end{eqnarray}
where $N_R$ is the total number of Right-electrons,
$[\varphi_{0,R}, N_R ]=i$ and $L$ is the length of the system.
$\alpha$ is a short-distance cutoff that must be introduced in
order to have converging integrals\cite{lut}. In principle the
bosonization provides exact results in the limit $\alpha \to 0$,
however for practical purposes it is useful to take a non-zero
(small) $\alpha$ which introduces a finite effective bandwidth
$\gamma = v_{F}/\alpha$ in the system. By doing this we have to
bare in mind that such procedure gives an accurate physical
description only in the low-energy part of the
spectrum\cite{voit1}.

The coupling of valence electrons to the core-hole is given
by\cite{gunn}\cite{lambda1}\cite{lambda2}
\begin{equation}
H_{\lambda}=\sqrt{\frac{2 \pi}{L}}\sum_{q \neq 0} \sum_{\sigma}
 \lambda(q) [b^{\dagger}_{\sigma}(q)+b_{\sigma}(q)] (1-n_{c}) \, ,
\end{equation}
where $\lambda(q)$ is the core-valence coupling constant, $L$ is
the volume of the system, $c^{(\dagger)}_{c}$ is the annihilation
(creation) operator of the core-\textit{electron}, whose occupancy
and energy are  $n_{c}=c^{\dagger}_{c}c_{c}$ and $\varepsilon_{c}$
respectively. In the following we will take $ \lambda (q) \equiv
\lambda $.

The term responsible for the Auger decay is more conveniently
expressed in the fermionic representation and reads:
\begin{equation}
H_{A}=c^{\dagger}_{p}c^{\dagger}_{c} \, A + \mathrm{h.c.} \, \, \,
, \, \, \,  A= V  \psi_{\uparrow}(0)\psi_{\downarrow}(0)
 \, ,
\end{equation}
where $c^{(\dagger)}_{p}$  destroys (creates) the Auger electron
and
\begin{equation}
 \psi_{\sigma}=\psi_{\sigma,R}+\psi_{\sigma,L}
 \, .
\end{equation}
$V$ is the so-called Auger matrix element, which here is
taken as a constant. Here we are assuming for simplicity that the
CVV decay leaves the two final holes in the origin of the system
in a singlet configuration. This reflects the local nature of the
Auger process; however such assumption is not essential and could
be relaxed.

As long as the interactions do not depend on spin, $H_{\rm{Lutt}}$
can be diagonalized  by introducing charge and spin boson
operators: $b^{(\dagger)}_{c,s}(q) =[b^{(\dagger)}_{\uparrow}(q)
\pm b^{(\dagger)}_{\downarrow}(q)]/\sqrt{2}$ and performing a
Bogoliubov transformation in the charge sector: $
\tilde{b}_c(q)=\cosh \varphi   b_c(q)+\sinh \varphi
b^{\dagger}_c(-q) $, $\tilde{b}^{\dagger}_c(q)=\sinh \varphi
b_c(-q)+\cosh \varphi  b^{\dagger}_c(q)$ with $ \tanh 2 \varphi =
(g_2/\pi)/(v_F+g_4/\pi) $ and renormalized velocity $v=[(v_F
+g_4/\pi)^2-(g_2/\pi)^2 ]^{1/2}$. In the next Section we use the
bosonization scheme sketched above compute the Auger spectrum of a
1D metal described within the Luttinger liquid theory.

\section{Calculation of the Auger spectrum}

The 1-step formulation of the the Auger processes has been
provided by GS\cite{gunn}\cite{gunn2} who showed that the Auger
current is given by the following
correlator\cite{gunn}\cite{cinidrchal}:
\begin{equation}
j(\omega)=\frac{\pi \alpha^{2}}{2}\int_{0}^{\infty} dt
\int_{0}^{\infty} dt' \, e^{i \omega(t-t')} \, f(t,t')
\label{gusc}
\end{equation}
where the factor $\pi \alpha^{2}/2$ is chosen in order to have a
normalized spectrum and
\begin{eqnarray}
 f(t,t')  \nonumber \\ =\langle g |  c^{\dagger}_{c}
e^{i(H[0]+i\hat{\Gamma})t'}   c_{c}  A^{\dagger} e^{iH[1](t-t')} A
c^{\dagger}_{c}  e^{-i(H[0]-i\hat{\Gamma})t} c_c |g\rangle .
\label{fttp}
\end{eqnarray}
In the above expression $|g\rangle$ is the ground state before the
X-ray photoemission, $H[0,1]$ is the Hamiltonian of the system
$H=H_{\mathrm{Lutt}} +\varepsilon_{c} (1-n_{c}) +H_{\lambda}$ with
$n_{c}=0,1$ respectively, and $\hat{\Gamma}$ is an effective
optical potential describing virtual Auger transitions and
relaxation processes\cite{gunn}.
In order to proceed we make the following approximation:
\begin{eqnarray}
f(t,t')& \approx & \langle \tilde{g}|  \, e^{i\tilde{H}[0]t'} \,
A^{\dagger} \,
e^{iH[1](t-t')} \, A  \, e^{-i\tilde{H}[0]t} \,|\tilde{g} \rangle \nonumber \\
& \times & e^{-i \varepsilon_{c} (t-t')}e^{-\Gamma (t+t')}
\nonumber \\
& \equiv & C(t,t') \times e^{-i \varepsilon_{c} (t-t')}e^{-\Gamma
(t+t')}  \label{assum}
\end{eqnarray}
where the second line of the above equation is the core-hole
Green's function with lifetime $1/\Gamma$ which is a c-number. $|
\tilde{g} \rangle$ denotes the ground state of $H_{\mathrm{Lutt}}$
(whose elementary excitations are created by
$\tilde{b}^{\dagger}_{c}$ and $b^{\dagger}_{s}$), describing the
valence band in the initial state, and $\tilde{H}[0] \equiv H[0] -
\varepsilon_{c}$. As noticed by GS\cite{gunn}, the strength of the
effective optical potential is proportional to the square of the
Auger matrix element, and hence we can replace  $V^{2}$ by
$\Gamma$. $C(t,t')$ can be calculated exactly by using the
bosonization formulas in Eqs.(\ref{bos1},\ref{boson}) and the
equations of motion method. After some algebra one gets a compact
expression by introducing new variables $\tau =t-t'$ and
$T=(t+t')/2$:
\begin{equation}
C(\tau,T)=\frac{\Gamma}{2\pi} \left[  \frac{\alpha^{g} e^{
h(\tau,T)}}{(-i\tau v + \alpha)^{g}} +  \frac{ \alpha^{l+1} e^{
k(\tau,T)}}{(-i\tau v + \alpha)^{l} (-i\tau v_{F} + \alpha)}
\right] \, , \label{ctt}
\end{equation}
where we have defined $g=2(\cosh^2 \varphi + \sinh^2 \varphi)$ and
and $l= (\cosh  \varphi+\sinh  \varphi)^{2}$. The complex
functions $h(\tau,T)$ and $k(\tau,T)$ are reported in Appendix A.
Finally the Auger current reads
\begin{equation}
j(\omega-\varepsilon_{c})= \int_{-\infty}^{\infty} d\tau
\int_{0}^{\infty} dT \, e^{i \omega \tau} \, e^{-2\Gamma T} \,
C(\tau,T) \, , \label{spec}
\end{equation}
where we refer the kinetic energy of the Auger electron $\omega$
with respect to the core-level energy $\varepsilon_{c}$.
Eqs.(\ref{ctt}) and (\ref{spec}) constitute the main finding of
the present work. In the next Section we discuss the most relevant
features emerging from Eqs.(\ref{ctt}) and (\ref{spec}), which
give a hint to understand the physics of the Auger transitions in
1D systems.

\section{Discussion}

\begin{figure}
\begin{center}
\mbox{\epsfxsize 9.5cm \epsfbox{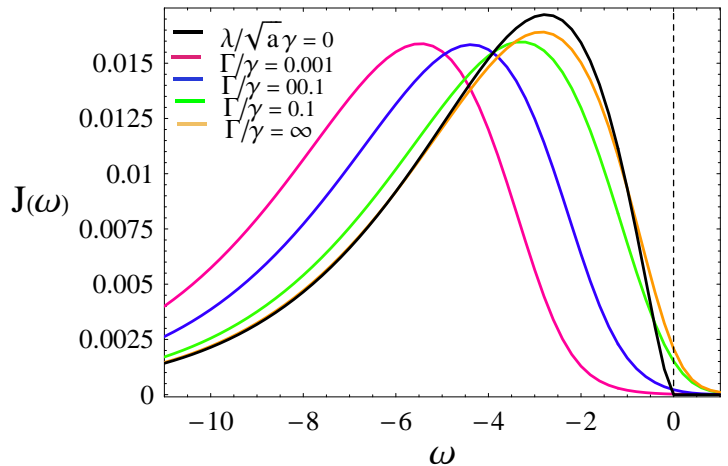}}
\end{center}
\caption{Auger current $j(\omega -\varepsilon_{c})$ calculated
numerically from Eq.(\ref{spec}) for different values of the
core-hole lifetime and core-valence interaction. Here we have
taken $g_{4}=g_{2}=4$, $\alpha =0.1$, $v_{F} =1$, $a=1$, and
$\lambda=4$ except in black curve, where $\lambda =0$. $\omega$ is
expressed in units of $\gamma$. The vertical dashed line denotes
the 2-step high-energy threshold $\varepsilon_{c}$, set equal to
zero in the present figure. } \label{1step}
\end{figure}

We first observe that despite the LL nature of the valence
electrons, the correlator $C(\tau,T)$ does \textit{not} obey a
power-law, which is spoiled by the interaction $\lambda$ between
the valence electrons and the core-hole. It is also interesting to
study the relationship of our solution with the 2-step approach.
This is done in the limit $\Gamma \to 0$. As discussed by GS, if
such limit exists, one should recover the well-known 2-step
solution since the Auger transition happens after the complete
relaxation of the initial state. Such limit is carried out  by
observing that
\begin{equation}
\lim_{\Gamma \to 0} 2\Gamma \int_{0}^{\infty} dT e^{-2\Gamma T}
e^{z(\tau,T)}=\lim_{T \to \infty }e^{z(\tau,T)} \, ,
\end{equation}
with $z=h,k$. We note that for any finite $\tilde{\lambda}$ the
limit on the r.h.s. does not exist because for large $T$ we have
$h(\tau,T) \sim k(\tau,T) \sim (\tilde{\lambda}/v)^{2} iv\tau \ln
(vT/a)$. This is a remarkable result, showing that the 2-step
approach is \textit{not} justified if the valence band is
described by the LL. On the other hand if we set
$\tilde{\lambda}=0$ the 1-step and 2-step solutions do coincide
because the 2-step spectrum is obtained from the 2-particle
Green's function describing the valence electrons (holes) in the
ground state of $H_{\mathrm{Lutt}}$. The 2-step approach is often
employed in typical Auger calculations and therefore it is
instructive to compare it with our 1-step solution. The 2-step
Auger current is readily obtained by setting $\tilde{\lambda}=0$
and results
\begin{eqnarray}
&& j_{2-\mathrm{step}}(\omega-\varepsilon_{c})  \nonumber \\
&=& \int_{-\infty}^{\infty} d\tau \frac{e^{i \omega \tau}}{4\pi}
\left[ \frac{\alpha^{g}}{(-i\tau v + \alpha)^{g}} + \frac{
\alpha^{l+1} }{(-i\tau v + \alpha)^{l} (-i\tau v_{F} + \alpha)}
\right]
 \, , \label{2step}
\end{eqnarray}
which recovers the characteristic power law suppression at $\omega
\approx 0$. The comparison between the Auger spectra calculated
with $j(\omega-\varepsilon_{c})$ and
$j_{2-\mathrm{step}}(\omega-\varepsilon_{c})$ is shown in
Fig.\ref{1step}. As discussed above, it is seen that $j$ does not
approach $j_{2-\mathrm{step}}$ for small $\Gamma$ (compared to
$\gamma$). In particular we note that for $\Gamma \to 0$ the
center-of-gravity $\varepsilon_{g}$ of $j$ (blue and violet
curves) is shifted towards lower kinetic energies with respect to
the center of gravity of $j_{2-\mathrm{step}}$ (black curve)  with
a logarithmic dependence $ \varepsilon_{g} \propto
\tilde{\lambda}^{2} \ln (\Gamma a/v)$.

Another interesting case is obtained in the limit $\Gamma \to
\infty$, that is for very short core-hole lifetime. In this case
$2\Gamma e^{-2\Gamma T}$ produces a Dirac delta in the
$T$-integration: \begin{equation}
 \lim_{\Gamma \to \infty}
2\Gamma \int_{0}^{\infty} dT e^{-2\Gamma T} e^{z(\tau,T)}= e^{
z(\tau,0)} \, ,
\end{equation}
with $z=h,k$. In this limit the Auger transition occurs when the
initial state is still excited, since
$e^{-i\tilde{H}[0]t}|\tilde{g} \rangle$ is not an eigenstate of
$H[1]$. As a consequence the excitations created on emission of
the initial core-electron transfer their energy to the Auger
electron, which then has a kinetic energy exceeding the
high-energy threshold $\varepsilon_{c}$ in the 2-step model [see
Eq.(\ref{2step}) and Fig.\ref{1step} (black curve)]. This
phenomenon, known as shake-down, is a typical example of
qualitative departures from the predictions of the 2-step model.

Finally, we show that our theory provides an explanation of the
sizable suppression of CVV Auger spectral weight closed to the
Fermi energy observed in carbon nanotubes with respect to
graphite\cite{dementjev}\cite{noi}. This is a striking trend,
since the structure of the one-particle density of states (1PDOS)
of the two carbon structures would predict the opposite behavior.
In fact, while in metallic nanotubes the 1PDOS  at the Fermi
energy is \textit{finite} due to the 1D linear dispersion, in
graphite it is \textit{vanishing}, due to the 2D conical
dispersion. Therefore we expect that correlation effects have to
be invoked in order to revert the one-particle scenario.

Metallic carbon nanotubes are believed to be rather good (although
approximate) realizations of LL\cite{gao}\cite{yao}\cite{bock}
since in normal conditions the main correlation effects come from
the long-range part of the Coulomb repulsion. In nanotubes with
radius $R$, the back-scattering interactions with large momentum
transfer suffer a $1/R$ suppression\cite{eg}\cite{kane}. Therefore
we believe that typical metallic (10,10) nanotubes are well
described by the present theory. Concerning graphite, we use the
CS approach, which is known to give the Auger spectrum in
excellent agreement with experimental one\cite{noi}\cite{houston}.
In order to employ the CS approach we must compute the 2-particle
valence Green's function within the bare ladder
approximation\cite{cini1}\cite{saw}. This is accomplished starting
from the non-interacting valence 1PDOS \begin{equation}
\rho_{0}^{2D}(\omega)= \gamma^{-2} \theta(-\omega) |\omega|
e^{-|\omega|/\gamma} \, ,
\end{equation}
which is obtained by imposing the 2D linear spectrum
$\varepsilon(k_{x},k_{y})=v_{F} (k_{x}^{2} +k_{y}^{2})^{1/2}$ and
the momentum cutoff $1/\alpha$. We note en passing that
$\rho_{0}^{2D}$ vanishes linearly at $\omega=0$, as it should. The
corresponding non-interacting 2-particle Green's function
$G^{2D}_{0}$ is obtained by self-folding $\rho_{0}^{2D}$ and by
Hilbert transforming:
\begin{eqnarray}
G^{2D}_{0}(\omega)&=&(1/ 6\gamma ) [2-(\omega /\gamma) +2 (\omega
/ \gamma )^{2} \nonumber \\
 &-&( \omega /\gamma)^{3}e^{\omega
/\gamma} \Gamma(0,\omega /\gamma)] \, ,
\end{eqnarray}
where $\Gamma(x,y)$ is the incomplete gamma function. Thus the
Auger spectrum of graphite according to CS theory is\cite{nota3}
\begin{equation}
j_{\mathrm{CS}}^{2D}(\omega-\varepsilon_{c})= -\frac{1}{\pi} \,
\mathrm{Im} \left[ \frac{G^{2D}_{0}(-\omega+i0^{+})}{1-U \,
G^{2D}_{0}(-\omega +i0^{+})} \right] \, ,\label{cscurr}
\end{equation}
where $U$ is the short-range screened repulsion felt by the two
valence holes in the final state. It is worth to recall that our
model is suitable to represent the $\pi$ electrons of nanotubes
and graphite, which are the ones involved in proximity of the
Fermi level (placed at $\omega=0$). Therefore only the low-energy
portion of the experimental Auger spectra can be addressed within
the present framework, while the high-energy spectral region,
corresponding to deep $\sigma_{p}$ and $\sigma_{s}$ states, cannot
be described here. A complete analysis including the missing
$\sigma_{p}$ and $\sigma_{s}$ states can be found in
Ref.\cite{noi}. However, in that paper the suppression of the
Auger spectrum of nanotubes closed to the Fermi energy was
reproduced by including some phenomenological form factors which
have been fitted with the experimental data. Conversely in the
present work the problem is treated starting form a fully
microscopic theory with no adjustable parameter.

\begin{figure}
\begin{center}
\mbox{\epsfxsize 8.5cm \epsfbox{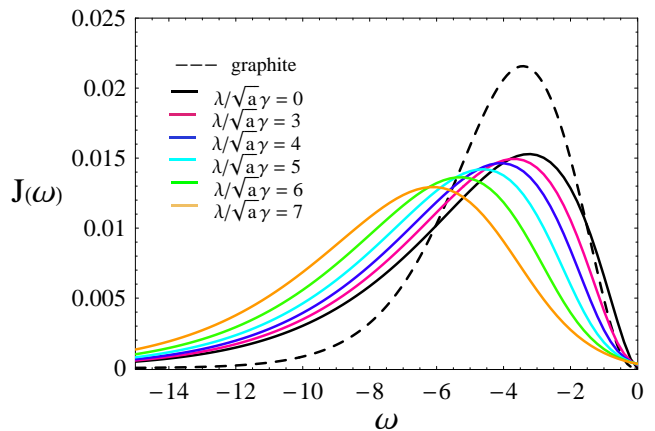}}
\end{center}
\caption{Comparison between the calculated Auger current for
graphite $j_{\mathrm{CS}}^{2D}(\omega-\varepsilon_{c})$ (dashed
black line) and  (10,10) SWNTs $j(\omega-\varepsilon_{c})$
according to Eq.(\ref{spec}) (solid lines) for different values of
the core-valence interaction. $\omega$ is expressed in units of
$\gamma$ and $\varepsilon_{c}$ is set equal to zero. }
\label{nangrap}
\end{figure}

In order to compare with the experiment, we use the following
realistic values for graphite and (10,10) metallic nanotubes:
$v_{F}\approx 10^{6}$m/s, $\alpha$ such that $\gamma =
v_{F}/\alpha \approx 10$eV, $\Gamma \approx 0.2$eV\cite{core}
(i.e. $\Gamma/\gamma =0.02$), $g_{4}=g_{2}\approx 2e^{2}\ln
(L/R)/\kappa \approx 5 v_{F}$\cite{kappa}\cite{onoratoprl}, $a
\approx 1$\AA$\; \;$, $U \approx 2$eV\cite{noi}\cite{lof}, and
leaving the adimensional ratio $\lambda/\sqrt{a}\gamma$ as free
parameter. In Fig.\ref{nangrap} we see that the inclusion of e-e
correlations in carbon nanotubes according to the LL theory within
the 2-step approach (black solid line) is not enough to reproduce
the suppression of $j$ vs $j_{\mathrm{CS}}^{2D}$ closed to $\omega
\sim 0$.  On the other hand the full 1-step theory with finite
$\lambda /\sqrt{a} \gtrsim \gamma$ provides results in qualitative
agreement with the experimental trend\cite{notagraph}. This is a
quite reasonable finding, considering that the core-valence
repulsion is larger than (but of the same order of) the
valence-valence repulsion.

\section{Summary and conclusions}

Traditional photoemission and inverse photoemission experiments
which probe one-particle dynamical responses provide a
well-established tool for the understanding of strongly correlated
1D systems. A great amount of theoretical work devoted to this
problem has been published in the past, enlightening the role of
the LL concept to explain several
features\cite{haldane,schulz,giamarchi,schonhammer,voit,meden}.

Surprisingly the study of the 1D Auger transitions, which are
related to the two-particle dynamics, has been only poorly
addressed. On the other hand the Auger spectroscopy is a powerful
experimental technique which permits the characterization of the
correlations in solids and therefore is of crucial importance in
the study of strongly correlated systems.

In the present work we have developed a dynamical theory of the
Auger processes in 1D metals described within the LL theory. Our
theory includes the finite core-hole lifetime, the valence-valence
and the core-valence interactions as well. A typical 1-step
feature is observed in the limit of small core-hole lifetime, in
which the valence electrons cannot relax before the Auger
transition, and the shake-down phenomenon occurs. Remarkably it is
shown that the 2-step approximation is not valid for any finite
core-valence interaction, which also spoils the low-energy
power-law behavior typically expected in the LL. Only for
vanishing core-valence interaction the power-law is recovered.
Finally we have shown that our 1-step theory is able to reproduce
the low-energy suppression of Auger spectral weight observed in
carbon nanotubes with respect to graphite.

\

The author kindly acknowledges M. Cini for helpful discussions.

\section{Appendix A: the functions $h(\tau,T)$ and $k(\tau,T)$ }

The final expression for $C(t,t')$ in Eq.(\ref{ctt}) has been
obtained by employing the bosonization formulas in
Eqs.(\ref{bos1},\ref{boson}) and the equations of motion method.
In order to perform the sum over $q$ we used that $q=2 \pi n/L $
and took the large-$L$ limit. When doing this, it is useful to set
$L = aN$ where $N$ is the number of sites of the 1D system and $a$
is the lattice constant, and send $N \to \infty$. The functions
$h(\tau,T)$ and $k(\tau,T)$ obtained in this way read
\begin{eqnarray}
h(\tau,T)&=& -\frac{\tilde{\lambda}}{v} 2i \sqrt{\pi}  (\cosh \varphi - \sinh \varphi)\nonumber \\
&\times& \left[ \sqrt{ 3\alpha +2iv(T+\frac{\tau}{2})} + \sqrt{ 3\alpha -2iv(T+\frac{\tau}{2})} \right. \nonumber \\
&-& \left. \sqrt{ 3\alpha +2iv(T-\frac{\tau}{2})} - \sqrt{ 3\alpha
-2iv(T-\frac{\tau}{2})} \right] \nonumber \\
&+& \left( \frac{\tilde{\lambda}}{v} \right)^{2} \left[ - 4\alpha
\ln(\frac{2\alpha}{a}) +2( 2\alpha -iv\tau)\ln(\frac{ 2\alpha
-iv\tau}{a}) \right. \nonumber \\
&+& \left. [\alpha -iv(T-\frac{\tau}{2})]\ln[\frac{\alpha
-iv(T-\frac{\tau}{2})}{a}] \right.
\nonumber\\
&-& \left. [\alpha +iv(T-\frac{\tau}{2})]\ln[\frac{\alpha
+iv(T-\frac{\tau}{2})}{a}] \right. \nonumber\\
&-& \left. [\alpha -iv(T+\frac{\tau}{2})]\ln[\frac{\alpha
-iv(T+\frac{\tau}{2})}{a}]\nonumber \right. \\
&+& \left. [\alpha +iv(T+\frac{\tau}{2})]\ln[\frac{\alpha
+iv(T+\frac{\tau}{2})}{a}] \right] \, ; \label{acca}
\end{eqnarray}

and

\begin{eqnarray}
k(\tau,T)&=& \left( \frac{\tilde{\lambda}}{v} \right)^{2} \left[ -
4\alpha \ln(\frac{2\alpha}{a}) +2( 2\alpha -iv\tau)\ln(\frac{
2\alpha
-iv\tau}{a}) \right. \nonumber \\
&+& \left. [\alpha -iv(T-\frac{\tau}{2})]\ln[\frac{\alpha
-iv(T-\frac{\tau}{2})}{a}] \right.
\nonumber\\
&-& \left. [\alpha +iv(T-\frac{\tau}{2})]\ln[\frac{\alpha
+iv(T-\frac{\tau}{2})}{a}] \right. \nonumber\\
&-& \left. [\alpha -iv(T+\frac{\tau}{2})]\ln[\frac{\alpha
-iv(T+\frac{\tau}{2})}{a}]\nonumber \right. \\
&+& \left. [\alpha +iv(T+\frac{\tau}{2})]\ln[\frac{\alpha
+iv(T+\frac{\tau}{2})}{a}] \right] \, , \label{kappa}
\end{eqnarray}
with $\tilde{\lambda}=\lambda \sqrt{2}(\cosh \varphi -\sinh
\varphi )$.


\begin{thebibliography}{99}

\bibitem{cinirew}
For a review, see e.g. C. Verdozzi, M. Cini and A. Marini, J.
Electron Spectroscopy and Rel. Phen. 117, \textbf{41} (2001).

\bibitem{gunn}
O. Gunnarsson and K. Sch\"{o}nhammer, Phys. Rev. B {\bf 22}, 3710
(1980).

\bibitem{cini1}
M. Cini, Sol. State Commun.  {\bf 24}, 681 (1977).

\bibitem{saw}
G.A. Sawatzky, Phys. Rev. Lett. {\bf 39}, 504 (1977).

\bibitem{bennett}
P.Bennett, J. C. Fuggle, F. Ulrich Hillebrecht, A. Lenselink and
G. A. Sawatzky, Phys. Rev. B \textbf{27}, 2194 (1983).

\bibitem{lof}
R. Lof, M. A. van Veenendaal, B. Koopmans, H. T. Jonkman and G. A.
Sawatzky, Phys. Rev. Lett. \textbf{68}, 3924 (1992).

\bibitem{maiti}
K.Maiti, D. D. Sarma, T. Mizokawa and A. Fujimori, Phys. Rev. B
\textbf{57},1572 (1998).

\bibitem{LdaU}
V. I. Anisimov, F. Aryasetiawan and A. I. Lichtenstein, J. Phys.:
Cond. Matt.  {\bf 9}, 767  (1997).

\bibitem{antibound}
G. Seibold, J. Lorenzana and F. Becca, arXiv:0706.1424
(unpublished).

\bibitem{haldane}
F.D.M. Haldane, J. Phys. C: Solid State Phys. \textbf{14}, 2585
(1981).

\bibitem{schulz}
H.J. Schulz, J. Phys.: Sol. St. Phys.  {\bf 16}, 6769  (1983).

\bibitem{giamarchi}
T. Giamarchi and H. J. Schulz, Phys. Rev. B \textbf{39}, 4620
(1989).

\bibitem{schonhammer}
K. Sch\"{o}nhammer and V. Meden, Phys. Rev. B \textbf{47}, 16205
(1993).

\bibitem{voit} J. Voit, Rep. Prog. Phys. {\bf 58}, 977
(1995).

\bibitem{meden}
V. Meden, Phys. Rev. B \textbf{60}, 4571 (1999).

\bibitem{gdsv}
J. Gonz\`{a}lez, M. A. Mart\'{i}n-Delgado, G. Sierra and M. A. H.
Vozmediano, \textit{Quantum Electron Liquids and High-$T_c$
Superconductivity}, Chap. 4, Springer-Verlag, Berlin (1995).

\bibitem{lut}
A. Luther and I. Peschel, Phys. Rev. B {\bf 9} 2911 (1974).

\bibitem{voit1}
J. Voit, J. Phys.: Cond. Matt.  {\bf 5}, 8305  (1993).

\bibitem{lambda1}
K.D. Schotte and U. Schotte, Phys. Rev. \textbf{182}, 479 (1969).

\bibitem{lambda2}
K.D. Schotte and U. Schotte, Phys. Rev. \textbf{185}, 509 (1969).

\bibitem{gunn2}
O. Gunnarsson and K. Sch\"{o}nhammer, Surf. Sci. {\bf 89}, 575
(1979).

\bibitem{cinidrchal}
V. Drchal and M. Cini, J. Phys. C: Condens. Matter \textbf{6},
8549 (1994).

\bibitem{dementjev}
A. P. Dementjev, K. I. Maslakov, A. V. Naumkin, Appl. Surf. Sci.
{\bf 245}, 128 (2005).

\bibitem{noi}
E. Perfetto,  M. Cini, S. Ugenti, P. Castrucci, M. Scarselli, M.
De Crescenzi, F. Rosei and M. A. El Khakani, Phys. Rev. B
\textbf{76}, 233408 (2007).

\bibitem{gao}
B. Gao, A. Komnik, R. Egger, D. C. Glattli and A. Bachtold, Phys.
Rev. Lett. {\bf 92} 216804 (2004).

\bibitem{yao}
Z. Yao, H. W. Ch. Postma, L. Balents and C. Dekker, Nature {\bf
402}, 273 (1999).

\bibitem{bock}
M. Bockrath, D. H. Cobden, J. Lu, A. G. Rinzler, R. E. Smalley, L.
Balents and P. L. McEuen, Nature {\bf 397}, 598 (1999).

\bibitem{eg}
R. Egger and A. O. Gogolin, Phys. Rev. Lett. {\bf 79}, 5082
(1997); Eur. Phys. J. B {\bf 3}, 281 (1998).

\bibitem{kane}
C. Kane, L. Balents, and M. Fisher, Phys. Rev. Lett. {\bf 79},
5086 (1997).

\bibitem{houston}
J. E. Houston, J. W. Rogers, R. R. Rye, F. L. Hutson and D. E.
Remaker, Phys. Rev. B {\bf 34}, 1215 (1986).

\bibitem{nota3} In
Eq.(\ref{cscurr}) we use $G_{0}^{2D}(-\omega+i0^{+})$ instead of
$G_{0}^{2D}(\omega+i0^{+})$ because CS theory is formulated in
terms of binding energy while in the present paper we use the
kinetic energy scale.

\bibitem{core}
A. Goldoni, R. Larciprete, L. Gregoratti, B. Kaulich , M.
Kiskinova, Y. Zhang, H. Dai, L. Sangaletti, F. Parmigiani, Appl.
Phys. Lett. {\bf 80}, 2165 (2002).

\bibitem{kappa}
$\kappa \approx 2$ is the dielectric constant of carbon nanotubes,
see e.g. Ref.\cite{eg}.

\bibitem{onoratoprl} S. Bellucci, J. Gonz\`{a}lez and P. Onorato, Phys.
Rev. Lett {\bf 95}, 186403 (2005).

\bibitem{notagraph}
In order to compare the low-energy part of Fig.\ref{nangrap} with
the experiment reported in Ref.\cite{noi} one has to shift the
energy scale $\omega$ of 284.6eV, which is the $1s$ core-hole
binding energy in the two carbon structures.

\end{thebibliography}
\end{document}